\documentclass[useAMS,usenatbib]{mn2e}
\usepackage{psfig,rotating}
\usepackage{graphicx,color}
\usepackage{dcolumn}
\usepackage{natbib}
\DeclareMathAlphabet{\mathsc}{OT1}{cmr}{m}{sc}
\def\testbx{bx}%
\DeclareRobustCommand{\ion}[2]{%
\relax\ifmmode
\ifx\testbx\f@series
{\mathbf{#1\,\mathsc{#2}}}\else
{\mathrm{#1\,\mathsc{#2}}}\fi
\else\textup{#1\,{\mdseries\textsc{#2}}}%
\fi}

\def\h2{\ensuremath{\rm H_2}}

\def\kms{km\,s$^{-1}$}

\newcommand{\be}{\begin{equation}}
\newcommand{\en}{\end{equation}}

\def\zabs{$z_{\rm abs}$}
\def\zem{$z_{\rm em}$~}

\def\hi{H~{\sc i}~}

\def\kms{km~s$^{-1}$}
\title[H~{\sc i} and DIBs in an external galaxy at z = 0.08]
{Parsec-scale  structures and diffuse bands in a translucent interstellar medium at $z\simeq0.079$\thanks{Based on data obtained with 
EFOSC2 at the NTT of the European Southern Observatory 
(Prgm. ID: 085.A-0709(A); PI: Rahmani), VLBA (Prgm ID:BG208; PI: Gupta) 
and GMRT(Prgm. ID: 18-074 and 20-037, PI: Gupta) }}
\author[Srianand et al.] {R. Srianand$^{1}$\thanks{E-mail:anand@iucaa.ernet.in}, N. Gupta$^{2}$, H. Rahmani$^1$, E.  Momjian$^3$,  P. Petitjean$^4$ and 
\newauthor P. Noterdaeme$^4$  \\
$^{1}$ IUCAA, Postbag 4, Ganeshkhind, Pune 411007, India \\
$^{2}$ Netherlands Institute for Radio Astronomy  (ASTRON), Postbus 2, 7990 AA, Dwingeloo, The Netherlands\\
$^{3}$ National Radio Astronomy Observatory, 1003 Lopezville Road, Socorro, NM 87801, USA \\
$^{4}$ Institut d'Astrophysique de Paris,UPMC-CNRS, UMR7095, 98bis Boulevard Arago, 75014 Paris, France \\
}
\begin{document}

\date{Accepted. Received; in original form }

\pagerange{\pageref{firstpage}--\pageref{lastpage}} \pubyear{2011}

\maketitle

\label{firstpage}

\begin{abstract}
We present a detailed study of the QSO-galaxy pair [SDSS J163956.35+112758.7
($z_q = 0.993$) and SDSS J163956.38+112802.1 ($z_g = 0.079$)] based on observations carried out using 
the Giant Meterwave Radio Telescope (GMRT), the Very Large Baseline Array (VLBA), 
the Sloan Digital Sky Survey (SDSS) and the ESO New Technology Telescope (NTT). 
We show that the interstellar medium of the galaxy probed by the QSO line of sight has 
near-solar metallicity (12+log(O/H) = 8.47$\pm$0.25) and dust 
extinction (E(B-V)$\sim$ 0.83$\pm$0.11) typical of what is usually seen in 
translucent clouds. We report the detection of absorption in the $\lambda 6284$
diffuse interstellar band (DIB) with a rest equivalent width of 
1.45$\pm$0.20\AA. 
Our GMRT spectrum shows a strong 21-cm absorption at the redshift of the galaxy with an integrated
optical depth of 15.70$\pm$0.13 \kms. Follow-up VLBA observations show that
the background radio source is resolved into three components with a 
maximum projected separation of 89 pc at the redshift of the galaxy.
One of these components is too weak to provide useful 21-cm H~{\sc i} information. 
The integrated H~{\sc i} optical depth towards the other two components are higher 
than that measured in our GMRT spectrum and differ by a factor 2. By comparing the
GMRT and VLBA spectra we show the presence of structures in the 
21-cm optical depth on parsec scales. We discuss the implications
of such structures for the spin-temperature measurements in high-$z$ damped 
Lyman-$\alpha$ systems. The analysis presented here
suggests that this QSO-galaxy pair is an ideal target for studying the DIBs and
molecular species using future observations in optical and radio wavebands.
\end{abstract} 
%

\section{Introduction}

Understanding parsec scale H~{\sc i} opacity fluctuations in the 
interstellar medium (ISM) of galaxies and how they depend on the
feedback from in situ star formation is very important.
In the Galaxy, the radio observations of H~{\sc i} 21-cm absorption
towards high-velocity pulsars and extended radio sources, 
the optical observations of Na~{\sc i} absorption lines towards 
globular clusters and binary stars have shown that the diffuse ISM 
is structured on 
parsec to sub-parsec ($\sim$10AU) scales
\citep{Frail94, Bates95, Heiles97,Deshpande00, Rollinde03,Boisse05,Brogan05,Roy12}.
These small-scale structures, 
due to large over-pressures, cannot survive rapid evaporation 
in the standard pressure-equilibrium based models \citep[see for example,][]{McKee77} 
and therefore raises many important questions regarding the nature of the ISM.
It is possible that these are transient phases controlled by 
turbulence \citep{Maclow04}. 

Physical conditions in \hi gas associated with high redshift galaxies can be 
probed by 21-cm absorption. In particular, the  H~{\sc i} 21-cm optical depth
combined with a determination of $N$(H~{\sc i}) from Lyman$-\alpha$ absorption
yield an estimate of the spin temperature, $T_S$,
\citep{Kanekar03} which is a good
indicator of the kinetic temperature \citep{Roy06}.
However, to fully interpret high redshift 21-cm 
spin-temperature measurements one needs to know the parsec scale
structure of the H~{\sc i} gas. In turn,
if radio emission of the background QSO has 
structures at different scales then Very Large Baseline Interferometry (VLBI) 
spectroscopy can be used to probe the spatial variations of the
\hi gas opacity.

Due to limitations of existing facilities, VLBI spectroscopic observations 
of QSO absorbers are limited to low-redshifts only.
In the case of the \zabs = 0.0912 DLA towards B0738+313, \citet{Lane00}
found the background source to be partially resolved at milliarcsec 
(mas) scales. Within the measurement uncertainties they do not find any strong
variations in the H~{\sc i} optical depth across 20 pc. The
associated galaxy could be a low surface brightness dwarf at
an impact parameter of $\le$3.5 kpc without any signature of
ongoing star formation \citep[see][]{Turnshek01}. In the case
of the $z = 0.03321$ galaxy towards J104257.58+074850.5, the background
radio source is unresolved in the Very Large Baseline Array (VLBA) 
image and the 21-cm
absorption is found to be similar over 27.1$\times$13.9 pc
\citep[see][for details]{Borthakur10}. Note that the impact
parameter in this case is 1.7 kpc and the foreground galaxy is a
low luminosity spiral with low gas phase metallicity and  
dust extinction along the QSO sight line.

Alternatively, time variability of 21-cm absorption is seen in two
cases [the absorbers towards B0235+164 \citep{Wolfe82} and B1127-145 
\citep{Kanekar01}] suggesting that the H~{\sc i} gas is patchy 
on pc scales. Interestingly, the \zabs = 0.524 absorption system 
towards B0235+164 is one of the extra-galactic systems that shows 
Diffuse Interstellar Bands \citep[DIBs;][]{Junkkarinen04}. 
Using a large sample of  $z\le1.5$ Mg~{\sc ii} absorbers searched for 21-cm 
absorption together with mas scale radio images \citet{Gupta12}, 
have concluded that the H~{\sc i} gas is patchy with a typical 
correlation length of 30$-$100 pc based on different correlation analyses.

It has been shown that, like 21-cm absorption, one can use \h2 absorption 
to probe the cold neutral medium (CNM) 
in $z\ge1.8$ damped Lyman$-\alpha$ systems 
\citep[DLAs;][]{Ledoux03,Srianand05,Noterdaeme08}. Recently \citet{Balashev11}, using 
partial coverage arguments, derived the linear size of the \h2-bearing core and the 
H~{\sc i} envelope to be $0.15^{+0.05}_{-0.05}$ pc and $8.2^{+6.5}_{-4.1}$ pc 
respectively for the \zabs = 2.3377 DLA towards Q1232+082. \citet{Srianand12}, 
combining 21-cm and H$_2$ information for a sample of high-$z$ DLAs, 
concluded that the typical size of \h2 bearing clouds
is $\le15$ pc.

All this indicates that the ISM of high-$z$ galaxies is structured
on parsec scales. 
However, most of the evidences are based on indirect arguments. 
In addition one would like to connect the structures seen in H~{\sc i}
to other observable properties of the galaxy such as
metallicity, dust content, local star formation etc. 
We are performing a systematic search for 21-cm and OH absorption 
in a well selected sample of Quasar-galaxy pairs (QGP, a fortuitous 
alignment of a foreground visible galaxy with a distant 
background quasar)  with 
impact parameter $b\le20$\,kpc. In case of detection we also perform
optical long-slit and VLBA spectroscopic follow-up observations.
Initial results from the GMRT observations of 5 QGPs, with the redshift 
of the galaxies in the range 0.03$\le z_g \le 0.18$ are presented in 
\citet{Gupta10}.  

\begin{figure*}
\hbox{
\includegraphics[height=8cm,angle=0]{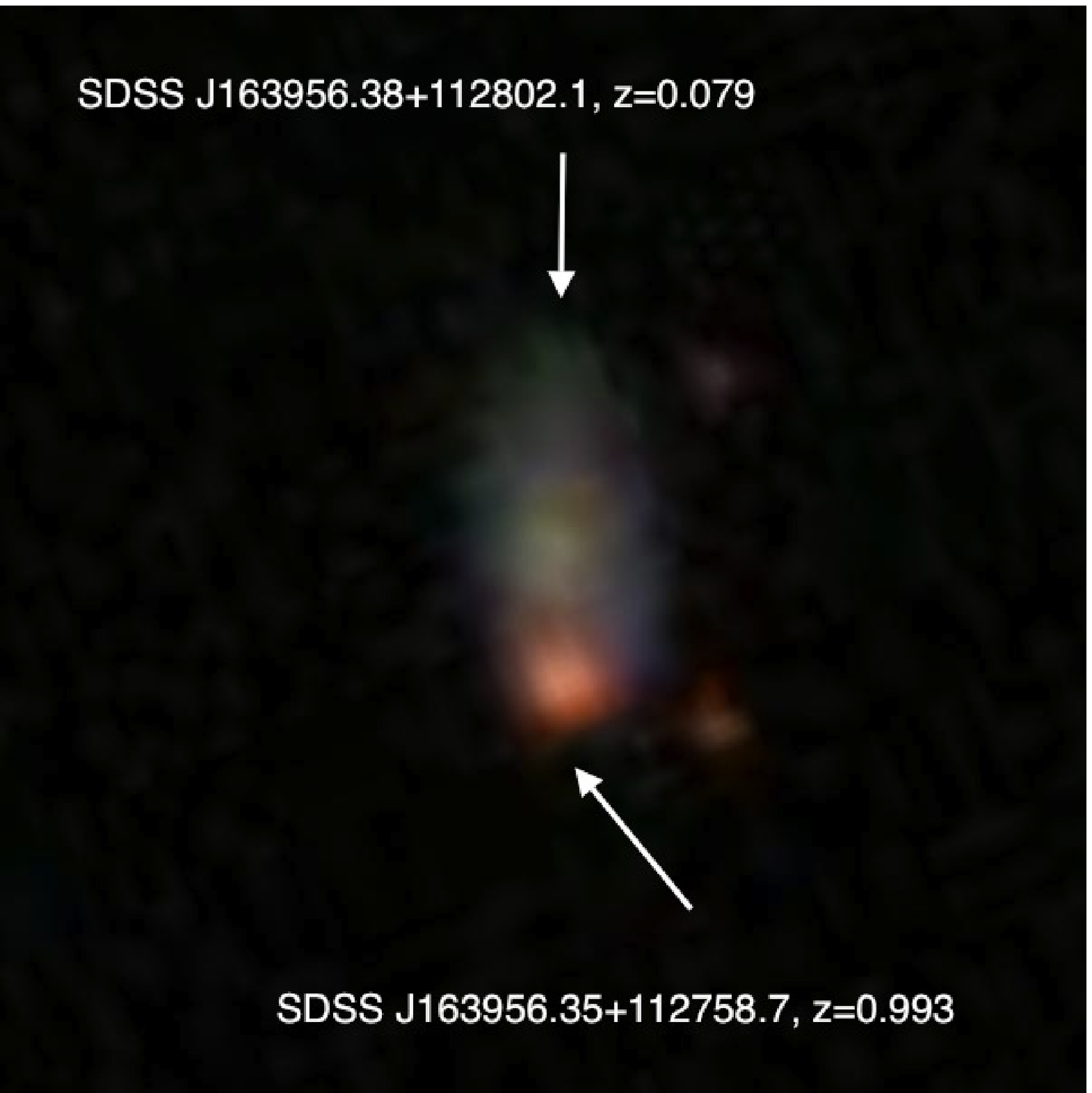} 
\includegraphics[bb=30 161 570 690,height=8cm,width=10cm,angle=0]{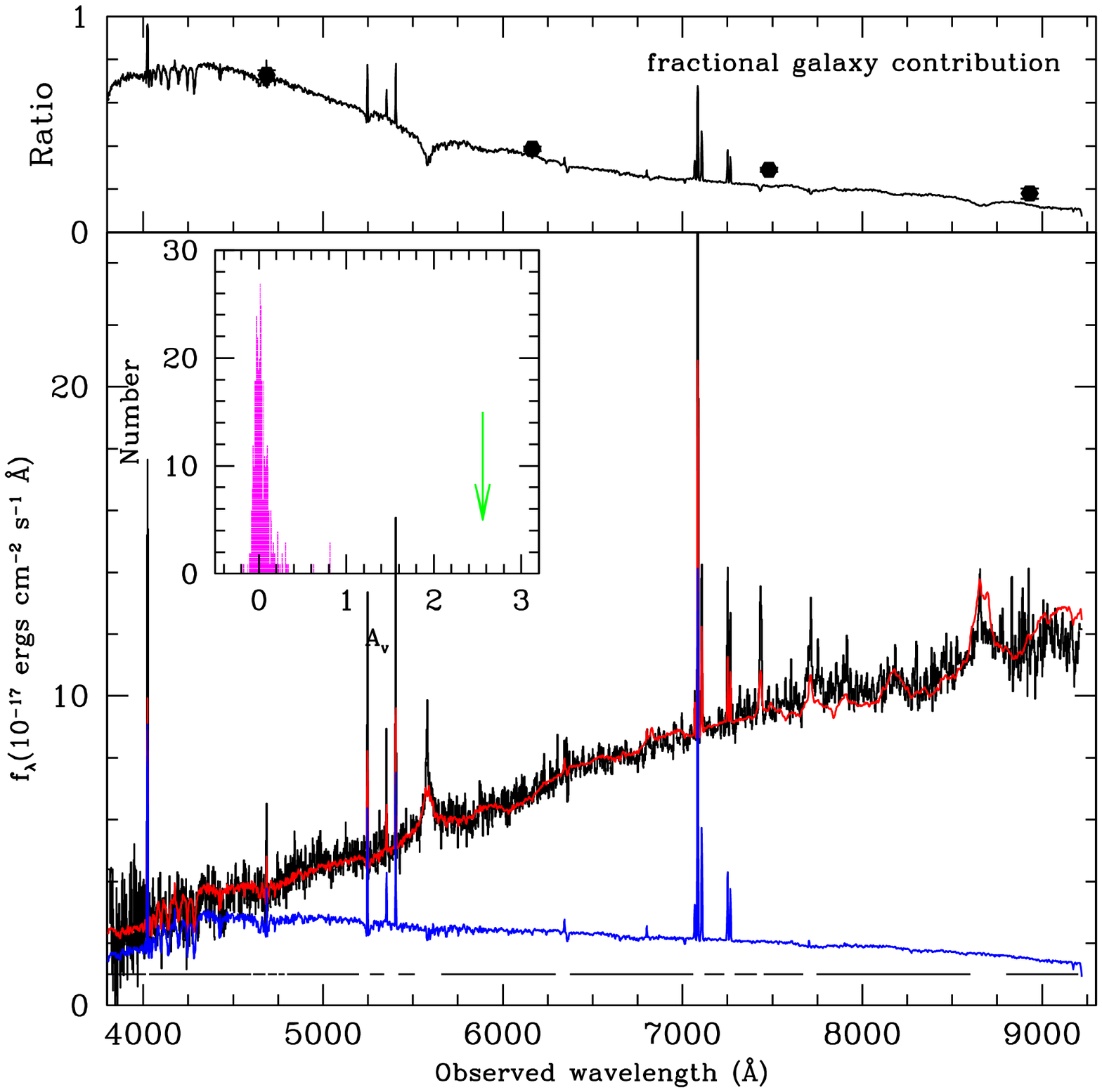} 
}
\caption{{\it Left:} SDSS colour image (30''$\times$ 30'') 
showing the QSO 
(SDSS J163956.35+112758.7, $z_q = 0.993$)  sight line piercing through 
the disk/spiral-arm of the foreground galaxy (SDSS J163956.35+112802.1, $z_g$=0.079)
{\it Right:} The SDSS spectrum of the QSO 
is fitted with a combination of a reddened QSO composite spectrum and a spiral
galaxy template (blue spectrum) from SDSS. The insert gives the 
distribution of A$_{\rm V}$ measured for a control sample of QSOs
with similar redshift as Q1639+1127. The arrow marks the 
measured A$_{\rm V}$ in the present case.
In the top right panel we show the relative contribution of the galaxy 
light to the SDSS spectrum as a function of wavelength. 
This contribution derived in different bands from SDSS images are also plotted 
as big black points.
}
\label{fig1}
\end{figure*}

Here we present a detailed analysis of a very special
QGP with the background QSO (J163956.35+112756.7, $z_q$ = 0.993) 
sight line piercing through the optical disk (Fig.~\ref{fig1}) of 
the foreground galaxy
(J163956.38+112802.1, $z_g$ = 0.079 or corresponding 
heliocentric velocity of $v_{\rm helio}\sim~23710$ \kms). 
We report the detection of 21-cm, Na~{\sc i} absorption and DIBs in this galaxy.
We use a long slit spectrum to study the star formation rate, metallicity
and rotation curve of the galaxy. The background radio source is
well resolved at mas scales. We present the VLBA spectroscopy and
discuss the parsec scale structures in the H~{\sc i} gas.
We will use a flat cosmological model
with $\Omega_m = 0.27$, h = 0.70 \citep{Komatsu09}.

\section{Physical conditions in the galaxy}

\begin{table}
\caption{Emission line parameters of the galaxy SDSS\,J163956.38+112802.1}
\begin{center}
\begin{tabular}{l c c }
\hline
\hline
 Line($l$) & F$_\lambda$(10$^{-17}$erg s$^{-1}$ cm$^{-2}$)&F$_\lambda$(H$\alpha$)/F$_\lambda$ \\
\hline
H$\alpha$                         & 147.0$\pm$2.9 & 1.00         \\
H$\beta$                          &  33.3$\pm$1.8 & 4.41$\pm$0.26\\
\mbox{[O~{\sc ii}]}$\lambda$3728  & 75.3$\pm$4.1  & 1.95$\pm$0.11\\
\mbox{[O~{\sc iii}]}$\lambda$4960 & 15.9$\pm$1.4  & 9.25$\pm$0.82\\
\mbox{[O~{\sc iii}]}$\lambda$5008 & 42.9$\pm$1.8  & 3.42$\pm$0.16\\
\mbox{[N~{\sc ii}]}$\lambda$6549  &  6.3$\pm$1.2  &23.17$\pm$4.55\\ 
\mbox{[N~{\sc ii}]}$\lambda$6585  & 28.7$\pm$1.8  & 5.11$\pm$0.34\\ 
\mbox{[S~{\sc ii}]}$\lambda$6718  & 28.5$\pm$1.7  & 5.16$\pm$0.33\\ 
\mbox{[S~{\sc ii}]}$\lambda$6732  & 20.3$\pm$1.5  & 7.24$\pm$0.55\\ 
\hline
\end{tabular}
\end{center}
\label{eline_t}
\end{table}
\subsection{Metallicity and dust in the emission line regions}
Strong emission lines from the foreground galaxy are seen in the SDSS 
spectrum (see Fig.~\ref{fig1}). They suggest the presence of a 
star forming region within a projected separation of 2.2 kpc to 
the QSO sightline\footnote{The SDSS fiber has a diameter of 3 arcsec and
was centered on the QSO. An angular scale of 1 arcsec corresponds to a
projected size of 1.474 kpc at the redshift of the galaxy. Therefore, the 
SDSS spectrum is the juxtaposition of the QSO spectrum with that of the galaxy within a radius of 2.2 kpc.}.
We fit the emission lines 
using Gaussians and the integrated line fluxes 
are summarised in Table~\ref{eline_t}.
A luminosity distance of 354.1 Mpc is used to convert these
fluxes into  luminosities.
The last column in this table gives the ratio of the H$\alpha$ flux
with respect to that of other emission lines.
Following Argence \& Lamareille (2009), we 
derive the optical depth in the systemic V-band of the galaxy, 
$\tau_V^{\rm Balmer}$ = 1.39$^{+0.18}_{-0.19}$, using
the observed H$\alpha$/H$\beta$ ratio, the intrinsic Balmer ratio of 2.85 
\citep{Osterbrock06} and the wavelength dependence of the
dust optical depth (i.e $\tau_\lambda$) as given by Eq. 3 of \citet{Wild07a}.
We estimate a dust uncorrected surface density of
the star formation rate (SFR) of 0.006 ${\rm M_\odot yr^{-1} kpc^{-2}}$. 
{Applying} the dust correction will increase this estimate by a factor of three for 
$\tau_V^{\rm Balmer}$ quoted above. 
If we assume that this disk galaxy obeys the  Kennicutt-Schmidt law then we get 
$N$(H~{\sc i})$\sim 2.6 \times 10^{21}$ cm$^{-2}$ using the
best fitted parameters of \citet{Kennicutt98,Kennicutt98b}.
Using the measured O3N2 ratio (i.e log [F$_\lambda$(O~{\sc iii}$\lambda$5007)/F$_\lambda$(H$\beta$) $\div$ 
F$_\lambda$([N~{\sc ii}]$\lambda$6585)/F$_\lambda$(H$\alpha$)]) and the best
fitted relationship given in \citet{Pettini04a} we get, 12+log~(O/H) = 8.47$\pm$0.25. Thus 
the QSO sightline is passing close to (i.e. within 2.2 kpc) a star forming 
region having near-solar metallicity and high $N$(H~{\sc i}). 

\subsection{Modelling the QSO  spectral energy distribution}

The background QSO spectrum is highly reddened (see Fig.~\ref{fig1}).
We get an independent estimate of $\tau_V$ and
E(B-V)  by fitting the 
QSO spectral energy distribution (SED) using, 
\begin{equation}
f_\lambda = A f_Q^{\lambda 1} e^{-\tau_{\lambda 2}} + B f_g^{\lambda 2} .
\end{equation}
Here $f_\lambda$ , $f_Q^{\lambda 1}$ and $f_g^{\lambda 2}$ are, respectively, the
observed flux at $\lambda$, the flux in the QSO composite
spectrum at $\lambda_1 = \lambda(1+z_q)$ and the flux in the galaxy continuum
that also enters the SDSS fiber. We approximate the latter with a spiral galaxy
template at $\lambda_2 = \lambda(1+z_g)$. 
A and B are normalization factors and $z_q$ and $z_g$ are the 
redshifts of the QSO and the foreground galaxy respectively. 
The form of $\tau_\lambda$ is taken from the  Milky Way extinction curve
\citep{Fitzpatrick88}.
The fitting method used is very similar to the one used in 
\citet{Srianand08} and \citet{Noterdaeme09co}. As can be seen from  
the right panel of Fig.~\ref{fig1}, 
the best fit (with $\chi^2_\nu = 1.23$) to the observed spectrum
is obtained for A$_{\rm V}$~=~2.56$\pm$0.02 (with E(B-V)= 0.82$\pm0.01$
and $R_V = 3.1$)\footnote{The derived value of E(B-V) depends on
our choice of $R_V$. In the absence of $R_V$ measurement we adopt $R_V = 3.1$ 
as seen in the Milky Way.}, 
{where A$_\lambda$ = 1.086 $\tau_\lambda$}. The quoted errors are 
mainly statistical ones. In Fig.~\ref{fig1} we 
also show the distribution of A$_V$ obtained for a control sample 
of SDSS QSOs with $z\sim z_q$.  {For this exercise we do
not consider the foreground galaxy contribution to the SED (i.e the 
second term in Eq.~1)}. The rms of the A$_V$ distribution of 0.11 
reflects a typical systematic error in the SED fitting method 
due to the dispersion of the QSO unreddened spectral energy distribution. 
Therefore, the reddening noted in the present case is significant at more than
the 20$\sigma$ level.
The measured value of A$_{\rm V}$ is consistent with the line of sight 
passing through a translucent region [defined as a region with 
1$\le$ A$_V \le $10 \citep[]{Vandishoeck89,Snow06}].
If we use the relationship between $N$(H~{\sc i}) and A$_{\rm V}$ derived in
the Galactic ISM \citep{Bohlin78} then the above inferred A$_{\rm V}$ is consistent
with $N$(H~{\sc i}) = $4\times 10^{21}$ cm$^{-2}$.
\begin{figure}
\vbox{
\includegraphics[bb=134 22 580 760,height= 8.5cm,angle=270,clip=true]{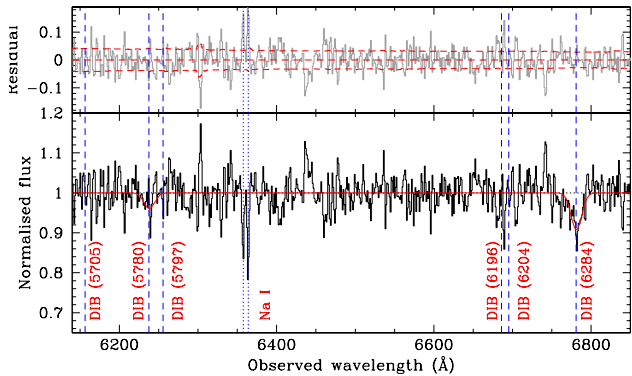} 
}
\caption{NTT spectrum of the QSO. 
Vertical dotted and dashed lines mark the positions of Na D lines
and DIBs respectively. A strong absorption is seen at the expected
position of the  $\lambda 6284$ DIB.  The best gaussian profile fitted to 
this feature is shown together with the absorption in the $\lambda 5780$~\AA\ DIB 
expected from the correlation found by \citet{Friedman11} in the Galactic ISM.
{\it Top-panel:} Residual flux after subtracting the Gaussian
fits from the observed spectrum. The 1$\sigma$ range
from the error spectrum (horizontal dashed line) is also shown. 
} 
\label{ntt1d}
\end{figure}
The difference between the two ${\rm A_V}$ estimates obtained using the Balmer decrement 
(i.e A$_{\rm V}^{\rm Balmer}~=~1.51^{+0.19}_{-0.18}$)  and the fit of the QSO
SED can be attributed to differences
in the dust opacity (or differences in $N$(H~{\sc i}), if we assume 
uniform dust-to-gas ratio) towards the QSO and the H$\alpha$ emitting
regions. 
In passing, we wish to point out that the line of sight discussed 
here has more dust opacity than any
of the high-$z$ dusty Mg~{\sc ii} and CO systems
that also produce 2175~\AA\ absorption
\citep{Srianand08,Noterdaeme09co,Jiang11}.

The SED fitting procedure used above predicts the relative contributions
of the QSO and the galaxy as a function of wavelength (top right panel in 
Fig.~\ref{fig1}). 
We independently estimate this ratio at different wavelengths using 
different SDSS broad band images within an aperture 
of 1.5" centered at the position of the quasar. We use the following steps: 
1) cut an image stamp of 80 pixels on a side around the 
QGP, 2) subtract a constant background obtained by averaging the counts 
in more than 10 neighboring regions close to the pair, 3) obtain the radial 
surface brightness profile of the galaxy (after masking the quasar) by fitting
isophotal ellipses  using  {STSDAS} package in IRAF, 4) derive the total counts 
due to the galaxy within an aperture of 1.5 arcsec radius at the location 
of the quasar using the fitted isophots.
The estimated ratios together with the errors (mainly from Poisson statistics) 
are also shown in the top-right panel of Fig.~\ref{fig1}. They are very much
consistent with the curve obtained from the SED fitting. This confirms
the robustness of our fitting procedure. In addition, based on
the ellipticity of the outer isophots in the $r$ and $i$ bands,
we measure a disk inclination angle of $\sim$ 63 deg.

\subsection{Na~{\sc i} absorption}
Using the European Southern Observatory (ESO) New Technology Telescope (NTT), 
we performed long-slit spectroscopic observations of
the QSO-galaxy pair aligning the slit (of width 1.2 arcsec { and
length 4.1 arcmin}) along the semi-major axis of the galaxy. The spectra (4$\times$2700s
exposures) were obtained in good seeing conditions ($\sim 0.8$ arcsec)
using the ESO Faint Object Spectrograph and Camera (EFOSC2) covering the 
wavelength range 6040$-$7135 \AA\ at a
spectral resolution of 200$-$250 \kms. The data were processed using
standard {\sc IRAF}\footnote{IRAF is distributed by the National Optical Astronomy Observatories,
which are operated by the Association of Universities for Research in Astronomy,
Inc., under cooperative agreement with the National Science Foundation.} long-slit routines. One dimensional spectra were
extracted with sub-apertures having a width  of 5 pixels (i.e projected 
size of $\sim$2 kpc). The results are summarized in Fig.~\ref{ntt1d}.

In the spectra of the QSO (shown in the bottom panel in Fig.~\ref{ntt1d}) 
we detect Na~{\sc i}$\lambda\lambda$5891,5897 
lines  with rest equivalent widths, $W_r = $ 0.65$\pm0.10$ and 
0.58$\pm$0.10\AA\ respectively at $z_g$. Using the empirical
relationship between E(B-V) and $W$(Na~{\sc i}) reported recently by
\citet{Poznanski12} we estimate E(B-V) = 0.31$^{+0.41}_{-0.12}$ and
0.47$^{+0.75}_{-0.29}$ from the measured equivalent widths. The
E(B-V) we measure from the SED fitting is higher than these
values but consistent within uncertainties.

As the doublet ratio is close
to 1 and the  resolution of our spectrum is low it is most likely
that the lines are saturated. 
A very conservative lower limit of log $N$(Na~{\sc i})$\ge$12.78
is obtained assuming the optically thin case. 
Consequently we derive $N$(H~{\sc i})$\ge 10^{21}$ cm$^{-2}$ from the known correlation
between $N$(Na~{\sc i}) and $N$(H~{\sc i}) found in our Galaxy 
\citep{Ferlet85, Wakker00}.

\subsection{Diffuse Interstellar Bands}

All the optical observations presented above suggest that the gas
along the line of sight has $N$(H~{\sc i}) $\ge 10^{21}$ cm$^{-2}$, with 
near solar metallicity and high dust content. In such cases one can expect
to detect absorption from different molecular species and diffuse
interstellar bands{ \citep[][]{Sarre06}}. In  Fig.~\ref{ntt1d}
we have marked the expected positions of DIBs using vertical 
dashed lines. A clear absorption is detected at the expected
position of the $\lambda_r = 6284$ \AA~ DIB. Using a Gaussian fit
we measure a rest equivalent width of 1.46$\pm$0.21 \AA\
and FWHM = 14$\pm$2 \AA. The measured FWHM is 1.4 times larger 
than what is seen in nearby starburst galaxies \citep{Heckman00}
and roughly a factor 2.2 higher than what is seen in the Galactic
ISM towards HD 204827\citep{Hobbs08}. 
No other DIB absorption is detected at more than the 3$\sigma$ significance
level. We place a  3$\sigma$ upper limit of 0.66 \AA\ for the rest
equivalent width of the $\lambda5780$ DIB feature.

\begin{figure}
\includegraphics[height=8.75cm,angle=0]{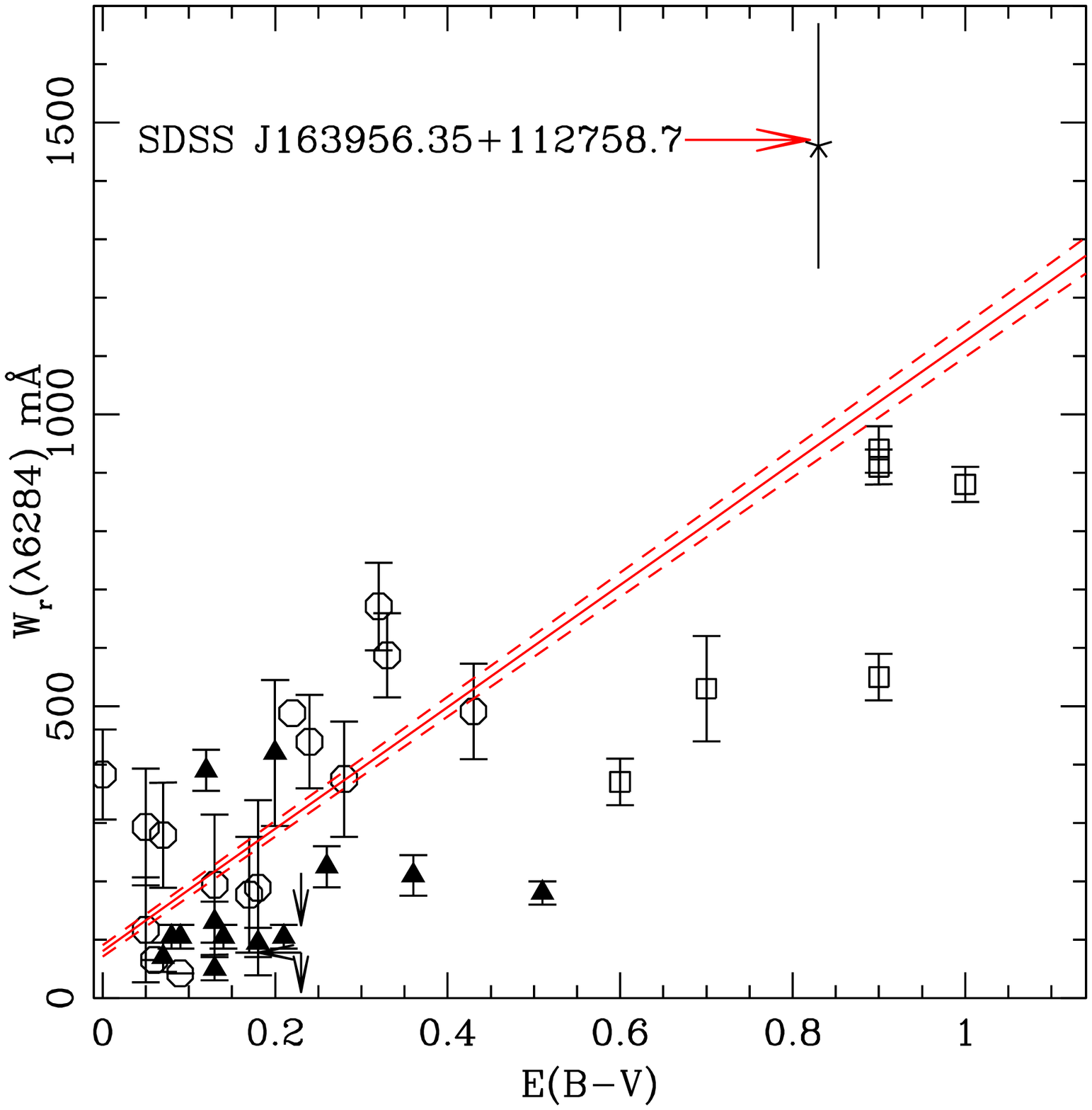} 
\caption{Comparison of E(B-V) and $W_r(\lambda6284)$
in different extra-galactic environments. The solid
and dashed lines are the best fitted relationship
and its 1 $\sigma$ error found in the milky way \citep{Friedman11}.
Filled triangles, open squares and open circles are for
Magellanic clouds \citep{Welty06}, nearby starburst galaxies
\citep{Heckman00} and M31 \citep{Cordiner11} respectively.
The data for the present system is marked with a star.
}
\label{dibstrength}
\end{figure}

In the local universe DIB observations are available for the ISM of
Magellanic clouds \citep[][]{Welty06,Cox06,Cox07}, M33 \citep[][]{Cordiner08},
M31 \citep[][]{Cordiner11}, {NGC 1448 \citep[]{Sollerman05}} and some nearby starburst galaxies 
\citep[][]{Heckman00}. In Fig.~\ref{dibstrength} we plot $W_r(\lambda 6284)$
against E(B-V) found in these cases. The strength of $\lambda 6284$
is affected by the strength of the radiation field and metallicity.
Low equivalent width is explained by  stronger 
radiation field and/or lower metallicity
It is interesting to note the metallicity measured in the present case
is close to that measured in M31 and starburst galaxies studied by
\citet{Heckman00}. However, the equivalent width of the $\lambda 6284$ DIB
is found to be higher than that predicted by the correlation found by
\citet{Friedman11}. This may imply slightly low ambient 
radiation field along the QSO sight line.  
Alternatively, if we assume that the physical conditions in the
galaxy studied here are similar to that of the Galactic ISM then
we get log~$N$(H~{\sc i}) = 21.76$\pm$0.25 and E(B-V) = 1.30$\pm$0.20
from the observed $W_r(\lambda 6284)$ and the correlations found
by \citet{Friedman11}. Interestingly the $N$(H~{\sc i}) value inferred here
matches well, albeit with large uncertainties, with that we infer 
from the $A_V$$-N$(H~{\sc i}) relationship of \citet{Bohlin78}. 
In addition, the inferred E(B-V) may mean $R_V$ = 1.9$^{+0.4}_{-0.2}$
and not 3.1 as we have assumed before.
This is similar to what has been inferred for the host galaxies 
of high-$z$ supernovae \citep{Wang06}.
Therefore, it is important to have independent measurements of $N$(H~{\sc i}) and $R_V$ for the present case.

In all the local measurements discussed above $W_r(\lambda5780)$ is found to be
smaller than $W_r(\lambda6284)$. If we use the correlations found by 
\citet{Friedman11}, we obtain $W_r(\lambda5780)$ = $0.61\pm0.08$~\AA. We 
show the expected profile in the bottom panel of  Fig.~\ref{ntt1d}.
From the residual plot shown in the top panel of Fig.~\ref{ntt1d} we can 
say that the observed spectrum is consistent with $\lambda5780$ DIB feature
having the predicted equivalent width.
From Table 3 of \citet{Welty06}, we find the
average value of the $W_r(\lambda6284)$/$W_r(\lambda5780)$ ratio
to be 2.9$\pm$0.2 in the case of the Magellanic clouds. Based on this 
we obtain $W_r(\lambda5780) = 0.5\pm0.1$ \AA.
Detecting this feature and other DIB features in a high signal-to-noise ratio
spectrum will allow us to probe the physical state of the gas in 
more detail.

This is only the third detection of DIBs due to
intervening absorbers in QSO spectra \citep[see][]{Junkkarinen04,Ellison08b}.
In previous cases the $\lambda 5780$ DIB was detected without
clear detection of the $\lambda 6284$ DIB. This is very much contrary
to what has been seen in the local universe \citep[apart from the rare sight
line towards Sk 143, as seen in][]{Welty06} and in the present case.  
\citet{York06}, by analogy with Sk143, attributed
these unusual line ratios to the ISM being  more protected from 
the ambient UV radiation field. 

The present NTT spectrum does not allow us to search for molecular
absorption. However, based on known correlations in the Galactic ISM
we expect log~$N$(CH)$\ge$13.8,  log~$N$(CH$^+$)$\ge$13.5 and
log~$N$(CN)$\ge$12.4 \citep{Welty06}. Therefore
high resolution follow-up spectroscopy of this source could yield
very good insights into the ISM of this external galaxy.

\subsection{H$\alpha$ rotation curve}
\begin{figure}
\vbox{
\includegraphics[bb=28 43 530 750,height=8.5cm,angle=90]{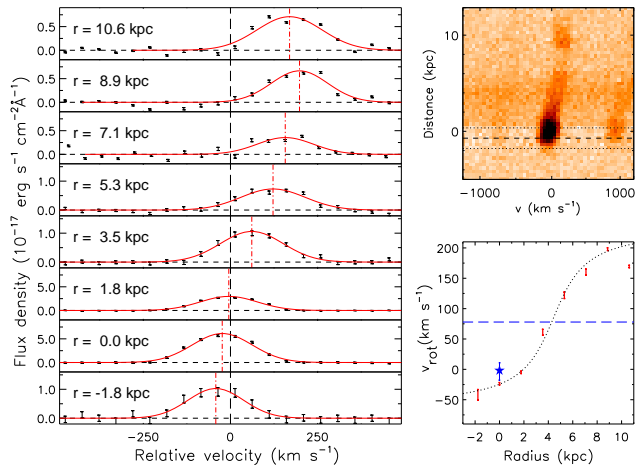} \\
}
\caption{{\it Left panel:}
Velocity plot showing the H$\alpha$ emission extracted at different
locations in the galaxy.  {\it Top right:} 2D spectra in the
region of H$\alpha$ and [N~{\sc ii}] emission lines after subtracting the
QSO light. The horizontal lines show the location of the QSO trace.
{\it Bottom right:} the rotation 
curve obtained from the H$\alpha$ line. The star is the velocity of the
H~{\sc i} gas seen in 21-cm absorption towards the QSO (see 
Section~\ref{gmrtsec}). The zero of the velocity scale
is set at \zabs = 0.0791 and zero of the spatial scale is set
at the position of the QSO.
} 
\label{ntt2d}
\end{figure}

In this sub-section we study the large scale kinematics of the emission
line gas using the NTT long-slit spectrum.
In the left panel of Fig~\ref{ntt2d} we show the H$\alpha$ emission
observed in different sub-apertures together with the best fitted Gaussian 
profiles. The measured FWHM is close to our spectral resolution suggesting
low velocity dispersion along the line of sight.
The 2D spectrum around the H$\alpha$ range after subtracting
the QSO trace is shown in the right panel. While the rotation
is apparent we also notice that the blue side is brighter 
suggesting additional star formation activity. Coincidentally the QSO
sightline is very close to this star forming region.
The angular separation between the center of the QSO trace and the center of 
the nearest star forming region (based on the peak of the H$\alpha$ 
emission) is 0.48 arc sec. This corresponds to a physical separation of 
0.7 kpc.  The different values of ${\rm A_V}$ measured toward 
the H$\alpha$ emitting region and the QSO reflect roughly a factor of 
2 change in ${\rm A_V}$ within a projected separation of 0.7 kpc. 
Therefore, 
the projected distance (and associated dust extinction) is large
enough so that the gas along the QSO sightline may not be influenced by
the star-forming region. The QSO sight line is at an impact parameter of 4 kpc
from the galactic center. The rotation curve plotted in Fig.~\ref{ntt2d}
suggests an asymptotic circular velocity of $\sim$ 125 \kms. 
This circular velocity for a typical radius of 10 kpc corresponds to a dynamical
mass of $\sim4\times10^{10}$ M$_\odot$ after correction for the inclination
angle of 63 deg. This suggests that the host galaxy is a low mass disk galaxy.

\section{21-cm absorption and parsec scale structures}


\begin{figure}
\includegraphics[height=8cm,angle=0]{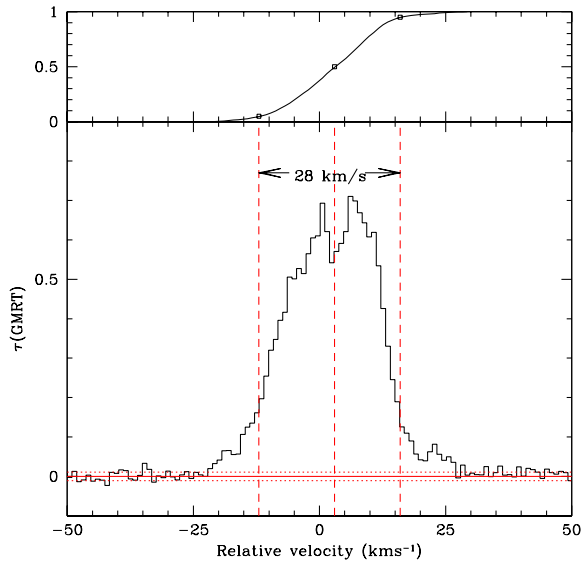} 
\caption{Optical depth profile of the 21-cm absorption towards 
 J1639+1127 detected in the GMRT spectrum. Top panel shows 
the cumulative optical depth profile. The open squares 
(and vertical dashed lines in the bottom panel)
mark the velocities at which the integrated optical depth
is 5\%, 50\% and 95\% of the total value. The mean optical
depth and associated RMS in the absorption free regions
are shown with solid and dotted horizontal lines respectively.
The velocity scale is defined with respect to $z = 0.079098$.
}
\label{gmrt21cm}
\end{figure}

\begin{figure}

\centerline{\includegraphics[height=8.5cm,angle=270]{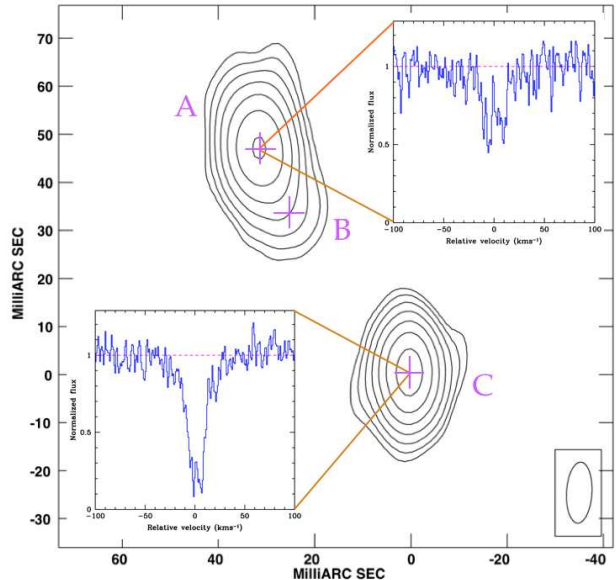} }
\caption{VLBA image showing J1639+1127 resolved into 3
Gaussian components. Crosses mark the locations of components A, B and C from top to bottom.
21-cm H~{\sc i} spectra at the peak of two strongest
components, A and C, separated by 89 pc, are also
shown. Absorption profiles are clearly different.
The RMS in the image is 8.425$\times 10^{-5}$ Jy/beam and the contour 
levels are 4.0$\times10^{-4}\times$(-2,-1,1,2,4,8,16,32, ...) Jy/beam.
The restoring beam (0.01271"$\times$0.00533" with a  position 
angle is -3.18 deg) is shown as an ellipse at the bottom of the figure.
Image center is at RA = 16 39 56.3624 and
DEC = +11 27 58.6568.
}
\label{figvlba}
\end{figure}

\begin{figure}
\includegraphics[height=8.5cm,angle=0]{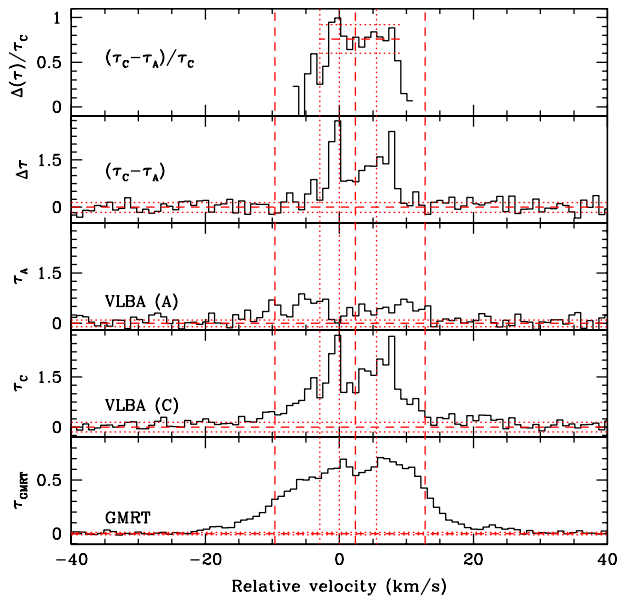} 
\caption{Bottom three panels show the 21-cm optical depth
profiles observed in, respectively, the VLBA spectra of components A, C, 
and the GMRT spectrum. The second panel
from the top shows the optical depth difference observed between
the two VLBA sightlines. The top panel shows the fractional change
in the optical depth between the two sight lines. The velocity
scale is defined with respect to $z= 0.079098$.  Vertical dashed
lines are as defined in Fig.~\ref{gmrt21cm} and the dotted lines
mark the velocities of narrow components seen towards C. }
\label{taufig}
\end{figure}

\subsection{GMRT 21-cm spectrum}\label{gmrtsec}
As the background QSO is radio-loud (with 1.4 MHz flux density of 
$\sim 164$ mJy) we get an opportunity to study the property of 
the cold H~{\sc i} gas along this dusty sightline. The source is
unresolved in the FIRST (Faint Images of Radio Sky at Twenty-centimeter) 
image with a deconvolved size of 1.32"$\times$0.82". As expected, a very 
strong 21-cm absorption was detected by Giant Metrewave 
Radio Telescope (GMRT) observations on 02 July 2010 
(2 MHz bandwidth split into 128 channels having a velocity resolution 
of $\sim$3.5 \kms per channel). We obtained a higher resolution 
(2.08 MHz bandwidth split into 512 channels having a velocity 
resolution of 0.93 \kms per channel) GMRT spectrum on 14 and 16 July 2011. 
All these data were reduced using the National Radio Astronomy 
Observatory (NRAO) Astronomical Image Processing System 
(AIPS) following the standard procedures. The radio source is
unresolved in our GMRT image (with a spatial resolution of 2.5"$\times$2.1") 
with a peak flux density of 155 mJy/beam.
The 21-cm optical depth profile from the high resolution spectrum
is shown in Fig.~\ref{gmrt21cm}. The integrated 21-cm optical 
depth is found to be, $\cal{T}=$15.7$\pm$0.13 \kms, with 90\% of it 
is within a velocity range of 28 \kms (see top panel in Fig.~\ref{gmrt21cm}). 
The observed $\cal{T}$ translates to $N$(H~{\sc i}) = $2.9\times 10^{19} (T_S/f_c)$ cm$^{-2}$ 
with $f_c$ being the covering factor, { the fraction of the background radio emission covered by the 
absorbing gas.} The probable values
of $N$(H~{\sc i}) discussed above based on known correlations suggest
that the harmonic mean spin temperature of the H~{\sc i} gas is 
less than or equal to 100 K as seen in the Galactic ISM.

Another interesting observation is that the H~{\sc i} gas that produces
the 21-cm absorption is redshifted with respect to the peak rotational 
velocity of the 
H$\alpha$ emitting gas at the same location by up to 25 \kms
(see Fig.~\ref{ntt2d}). 
This once again confirms that the absorbing gas and H$\alpha$ 
emitting gas are not co-spatial.

\subsection{Spatially resolved VLBA spectroscopy}

The VLBA observations of J1639+1127
were carried out on July 27 and October 1, 2011. The total and
on-source observing times were 10 and 6.3 hrs respectively. 
One 8 MHz baseband channel pair was used in these observations,
with right- and left-hand circular polarization sampled at
2 bits, and centered at the frequency of the redshifted H~{\sc i}
21-cm line. The data were correlated with 4096 channels (channel
width of 0.45 \kms) and 2 sec correlator integration time.
We followed the standard data reduction procedures  for reducing the 
VLBA data \citep[see for example][]{Momjian02,Srianand12}.
The radio continuum emission is resolved into multiple components at 
mas scales (Fig.~\ref{figvlba}).  
\begin{table}
\caption{Results of VLBA observations}
\begin{tabular}{cccccc}
\hline
\hline
Source   &\multicolumn{2}{c}{Flux density}  &  Deconvolved &  &\multicolumn{1}{c}{} \\   
             & Peak   & Total& angular size &PA &$\int \tau(v) dv$  \\
             &mJy/beam &  mJy & (mas$^2$) & (deg)& kms$^{-1}$\\
\hline
A          &    26.8   & 57.4    &9.9$\times$6.7 &  +31.6 & 17.6$\pm$1.6\\
B          &      1.7   &  2.1    &8.8$\times$0.0$^\dagger$ &  +107.2  &....\\  
C          &  36.8   & 56.8    &5.9$\times$4.1 &  +62.0& 37.6$\pm$2.9\\
\hline
\end{tabular}
\begin{flushleft}
$^\dagger$ An angular extent of 0 indicates a source size much smaller than the
synthesized beam.
\end{flushleft}
\label{tabvlba}
\end{table}
We use the multiple Gaussian fits to the VLBA images and found 3 distinct 
components. The results are summarised in Table~\ref{tabvlba}. 
Only 78\% of the flux density measured with the GMRT is 
recovered with the VLBA. 
Components A and C, that are separated by 55.8 mas ({implying 
a linear line of sight separation of 89 pc at the redshift of the galaxy}), have
nearly equal flux densities and contribute to 98\% of the total
flux density seen in the VLBA image. {The peak flux densities of
these components are 47\% and 65\% of their respective total 
flux densities, suggesting structures at few mas scales in the
radio emission}.

The 21-cm absorption toward the peak emission of A and C extracted from 
our VLBA observations (smoothed to a channel width of 0.9 \kms to 
match the GMRT spectrum) are shown in Fig~\ref{figvlba}. The 
integrated 21-cm optical depths, given in the last column of the 
Table~\ref{tabvlba}, are more than that measured in our GMRT spectrum and
differ by a factor of 2  within the 89 pc probed by the two sightlines. 
Unlike the GMRT spectrum, that is very smooth, the 21-cm spectrum
towards component C shows the presence of narrow absorption 
components superposed on a smooth absorption profile.
This confirms the patchy distribution of the absorbing gas at 
parsec scales. 

To probe this further, we compare in Fig.~\ref{taufig} the optical 
depth profiles seen in the GMRT and the VLBA observations. Fig.~\ref{taufig}
also shows the difference in optical depth (i.e $\Delta \tau$) 
{between the lines of sight towards A and C} as a function 
of velocity. While broad smooth absorption profiles with nearly
identical optical depths  are seen, narrow absorption 
components (identified with vertical dotted lines in Fig.~\ref{taufig}) 
are clearly seen only towards C. The top panel of Fig.~\ref{taufig}
shows the fractional difference in  $\tau$ between the two sight 
lines compared to $\tau_C$. It is clear that on average 
$\tau_C \sim 5\times \tau_A$ in the velocity range -5 to 10
\kms, where the optical depth towards C is dominated by several
narrow components. This basically confirms the large variations
in H~{\sc i} optical depth over a length scale of 89 pc for
narrow H~{\sc i} components.

\subsection{Tiny clouds and parsec scale structure}

As the deconvolved sizes of the VLBA
components correspond to a projected size of $\sim 8$ pc the next
question we wish to address is whether there is any optical depth
variation over this length scale. We do this by comparing the
optical depth profiles seen towards A and C with the total
optical depth profile seen in the GMRT spectrum. In general
we can write,
\begin{equation}
{\rm \tau_{\rm GMRT}(v) = f_A \tau_A(v) + f_C \tau_C(v) + \tau_{D}(v)}.
\end{equation}
Here, ${\rm f_A}$ and ${\rm f_C}$ are the covering factors of 
the gas component having optical depth ${\rm \tau_A(v)}$ and ${\rm \tau_C(v)}$
respectively. ${\rm \tau_{D}(v)}$ is the optical depth towards components
not seen in the VLBA image.
${\rm f_C}$ will be 0.37 if the absorbing gas covers all
the radio emission seen in the VLBA image from the component C
{(i.e ${\rm f_C}$ is ratio of flux density of component C to the
total flux density measured in the GMRT image).}
Therefore, if we assume a plane parallel gas slab covering only
the emission from component C then we expect ${\rm \tau_{\rm GMRT}(v) = 0.37 \tau_C(v)}$. 
If we assume ${\rm \tau_A(v) = \tau_C(v)}$, then we expect  ${\rm \tau_{\rm GMRT}(v) = 0.74 \tau_C(v)}$.  If   ${\rm \tau_A(v) \le \tau_C(v)}$ then we expect
${\rm 0.37 \le (\tau_{\rm GMRT}(v)/ \tau_C(v)}) \le 0.74$. On the contrary
if  ${\rm \tau_A(v) \ge \tau_C(v)}$ then 
${\rm \tau_{\rm GMRT}(v)/ \tau_C(v)} \ge 0.74$. Note this condition 
is also obtained when additional components are present towards the 
broad diffuse
emission resolved out in the VLBA image (i.e ${\rm \tau_D \ge 0 }$).
This discussion clearly suggests that by looking at the ratio of the optical
depth observed by GMRT and VLBA in one of the components we will be able to draw
broad conclusions about the optical depth variability at small scales. 

First we focus on the most interesting narrow component
at $v = 0$ \kms. This component is distinctly visible in C
and evident even in the GMRT spectrum however clearly 
absent towards A (as suggested by $\Delta\tau\sim \tau_C$).
The peak optical depth of this component is 2.5$\pm$0.1 
towards C. Based on the flux densities in the VLBA image,
if the absorbing gas covers only C then we expect the peak
optical depth at $v\sim 0$ \kms in the GMRT spectrum to be 
$\sim 0.93\pm0.05$. 
However, the measured value in the GMRT spectrum is $0.70\pm0.01$,
suggesting that, if the absorption component is a plane parallel
slab with uniform $N$(H~{\sc i}), then it would cover only 75\%  of the
continuum emission from component C. The observed parameters, on 
the contrary simply imply the presence of a strong optical depth
gradient within the deconvolved size of component C (i.e $\le$6 mas).
This angular scale corresponds to a  transverse size $<8.8$ pc. 
As this component is distinctly visible, we fit multiple 
Gaussian to $\tau_C$. While Gaussian fit to other components
may be unphysical, simultaneous fits allow us to get a 
realistic estimate of FWHM (1.75 \kms) and the peak optical
depth (2.5$\pm$0.1) for the narrow component. We get
an upper limit of 66 K for the kinetic temperature of the
gas from the FWHM. Using this as an indicator of spin
temperature ($T_S$), we get 
$N$(H~{\sc i}) = $5.4\times 10^{20} (T_s/66 K)$ cm$^{-2}$.
If we approximate the absorbing cloud as a sphere then we
get a particle density of $\sim$40 cm$^{-3}$
using $N$(H~{\sc i}) and the transverse size discussed above.
Physical conditions in this component are typical of 
diffuse molecular clouds.

\begin{figure}
\includegraphics[height=8cm,width=6cm,angle=270]{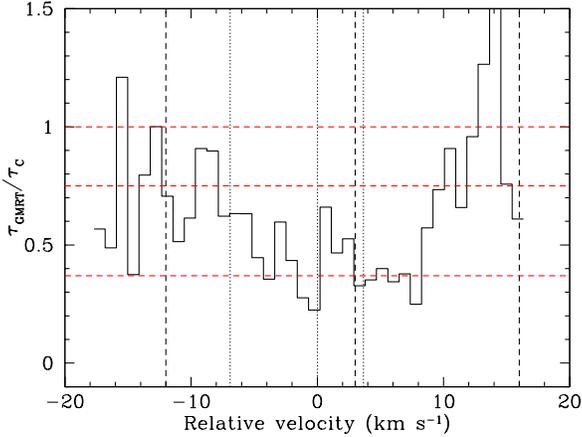} 
\caption{The ratio of $\tau_{\rm GMRT}$ and $\tau_{\rm C}$ as a 
function of relative velocity. The velocity scale and vertical
lines are as in Fig.~\ref{taufig}.}
\label{fcfig}
\end{figure}

To explore this further we plot in Fig.~\ref{fcfig} the ratio
of $\tau_{\rm GMRT}$ and $\tau_{\rm C}$ in the velocity range that
contains 90\% of the integrated optical depth. The expected
ratio when a parallel slab of gas with constant $\tau$ covers
only the component C,  all the VLBA components and all the flux 
seen by GMRT are shown by horizontal lines at 0.37, 0.74 and 
1 respectively. We use this plot to discuss the extent of the
gas that produces absorption in different velocity ranges.  

First we consider the velocity range $3 \le v$(\kms)$\le 9$ where
absorption is seen towards both A and C with $\tau_C \sim 5 \times \tau_A$
(see top panel in Fig.~\ref{taufig}). If the gas towards C and A 
covers all the radio emission from these components then we expect
$\tau_{\rm GMRT}/\tau_C$ to be 0.44 (i.e. $0.37\times(1.0+0.2)$). 
From Fig.~\ref{fcfig} we see the observed ratio is consistent
with 0.38 in this range. We notice that the optical depth error
in each channel in the VLBA spectrum towards C is $\sim$0.1.
For the mean optical depth measured in this velocity range 
this translate to an error of 0.03 in the ratio. Therefore,
in each channel we find the ratio to be lower than 0.44
by $2\sigma$ level. As this happens over 6 channels we 
see the difference to be significant at the 4.8 $\sigma$ level.
This difference can be understood if some of the narrow
components seen towards C have projected sizes less than 8.8 pc
(or the presence of strong opacity gradient within this scale)
as seen in the case of the narrow component at  $v\sim 0$ \kms.

In the velocity ranges  $10 \le v$(\kms)$\le 16$ and  
$-10 \le v$(\kms)$\le-4$ from Fig.\ref{taufig}, we find
$\Delta \tau \sim 0$. Therefore we expect the ratio of
$\tau_{\rm GMRT}$ and $\tau _C$  to be 0.74.
In the case of the first velocity range the ratio is above 0.74.
In the framework presented above, this could either mean 
$\tau_A\ge\tau_C$ or contribution to $\tau_{\rm GMRT}$ from the 
diffuse component resolved in the VLBA image. 
If we use the average value (1.2) of the measured ratio in this 
velocity range we get $\tau_A= 2.1\times\tau_C$.  
In the second velocity interval the ratio is found to be between 0.36 and 0.74. 
This is consistent with $\tau_A\le\tau_C$. Using the mean value of
the ratio (0.65) in this velocity range we find  
$\tau_c\sim1.4\tau_A$. These are very different from $\tau_C = \tau_A$ 
we see towards the peak emission in A and C.
Therefore, we can conclude that there are strong opacity
gradients at the spatial scale of $\sim$ 10 pc even in the
gas that produces absorption in this velocity range.

Finally in the velocity range -18 to -10 \kms, 
the absorption profile is smooth both in the VLBA spectrum
towards C and the GMRT spectrum. Therefore we smooth the spectra to
a 4 \kms resolution to increase the signal to noise ratio.
We notice that $\tau_{\rm GMRT}/\tau_C$ in this velocity range is
0.67$\pm$0.03. This just differs from the expected value of 0.74
by 2.3$\sigma$ level. Within the observational uncertainty
the optical depth seen towards A and C are nearly equal as
suggested by the difference in $\tau$ plotted in Fig.~\ref{taufig}.
Increasing the signal to noise of our VLBA spectra will help
detecting minor optical depth differences even in this smooth
component.

In summary, we find the 21-cm absorption has both
narrow and broad components. The narrow components 
show optical depth variations by a factor 1.4 to 10 over 
a 89 pc scale. In one of the narrow components we show that optical
depth variations are present over a 8 pc scale. 
On the contrary the absorption seen 
in the broad wings  are consistent with nearly similar optical 
depths (i.e at 2.5$\sigma$ level) within measurement uncertainty. 
Thus a simple picture of the system could be that of several
cold, dense and small clouds with characteristic size $\le 10$ pc 
embedded in a smooth diffuse H~{\sc i} medium covering several tens of parsec.
This scenario  is similar to the ``blobby sheets" seen in the Galactic
ISM \citep{Heiles03}.  

\section{Discussion}
We report the detection of diffuse interstellar bands and 21-cm absorption
from the \zabs~=~0.079 galaxy SDSS J163956.38+112802.1 towards the 
\zem= 0.993 QSO J163956.35+112758.7. 
The QSO line of sight in this case is passing through the disk of a relatively
low mass ($\sim$4 $\times 10^{10}$ M$_\odot$), near solar metallicity spiral
galaxy at an impact parameter of $\sim$4 kpc. The region with highest star
formation surface density, inferred from $H\alpha$ emission, in this
galaxy is located within a projected distance of 0.7 kpc from the
QSO sightline.

The QSO is highly reddened with E(B-V)$\sim$0.83$\pm$0.11 (estimated from the 
SED fitting) along the line of sight. This suggests that the 
line of sight is passing through a translucent interstellar medium.
The measured E(B-V), rest equivalent widths of Na D lines and 
surface density of star formation favor $N$(H~{\sc i}) $\ge 10^{21}$ 
cm$^{-2}$ along this sight line.

We report the detection of a $\lambda$6284 DIB feature
in the QSO spectra at the redshift of the galaxy. The strength of this feature
per unit reddening is found to be higher than what is seen in the Milkyway.
This could be related to differences in the DIB carrier and/or to the 
ambient radiation field. We also find the FWHM of this feature to be
broader than what is typically seen in our Galaxy but close to what
is seen in starburst galaxies. 
Our spectrum is consistent with the presence of the $\lambda$5780 DIB 
with a rest equivalent width predicted from the observed $\lambda$6284 DIB feature
and the correlation reported by \citet{Friedman11}. Due to poor SNR,
no other DIB feature is detected in our spectrum. In order to 
probe the physical conditions and the nature of the DIB carriers
it would be important to obtain high signal-to-noise ratio spectrum of
the QSO.

Strong 21-cm absorption from the galaxy, spread over 28 \kms, 
is detected in our GMRT spectrum. Our follow-up VLBA 
observations reveal the presence
of two strong radio emitting components, separated by 89 pc,
that are used to understand the spatial variations of the 21-cm optical depth. 
First we notice that the integrated optical depths towards these two components 
differ by a factor of 2. We show that this difference is mainly dominated
by narrow components that are seen only towards one of the VLBA component
(i.e component C). In these narrow components the H~{\sc i} opacity
seems to change over $\sim 10$ pc scales. For one of these narrow 
components we show the kinetic temperature is $\le$ 66 K. As
we expect dust to be mainly associated with low temperature gas, if 
the other narrow components are also cold, then we can speculate that
the optical emission from the QSO is associated 
with the VLBA component C.

In high-$z$ 21-cm absorption line studies 
of DLAs and Mg~{\sc ii} systems, and in the absence of VLBI spectroscopy, one uses 
the `core' fraction measured from the mas scale continuum images to correct
the optical depths for partial coverage 
\citep[see for example,][]{Kanekar09vlba,Srianand12,Gupta12}.  
In case of DLAs, this allows one to measure the spin temperature
of the H~{\sc i} gas. There are three issues that affect this approach:
(i) Unambiguous identification of the `core'. If the milliarcsecond morphology 
of the radio emission is not simple, then VLBI observations are needed to 
identify the flat-spectrum `core' component.
(ii) The fact that the radio sight line towards the `core' may still be tracing a larger gas volume 
with respect to the optical sight line. (iii) The implicit assumption that a single absorbing 
cloud covers the `core' i.e. a single covering factor ($f_c$).

The third issue is mainly related to the patchiness of the absorbing gas on scales of
tens of pc, which is typically the resolution provided by VLBI observations, 
and can only be addressed via VLBI spectroscopy. In the case of J1639+1127 
discussed here
it is plausible to associate component C with the quasar sight line 
because it shows clear signatures of the presence of a large amount of cold gas. 
Under the assumption of a single cloud covering C we would have estimated 
$f_c$ as 0.37 (57/155). However, if some of the absorption components were 
to cover only the peak of C, then $f_c$ would be off by as much as a factor 
of 1.5 (56.8/36.8) from the above value.
Furthermore, the covering factor estimate could be dramatically off if 
the core is associated with component B, resulting in $f_c$ of 0.01, or 
with a component not detected in our VLBA image. This is a possibility
considering that the radio continuum morphology of J1639+1127 at high angular 
resolution (see Fig~\ref{figvlba}) resembles that of a compact symmetric
object \citep[CSO;][]{Conway02}.
Here, the two dominant radio sources, A and C, would be the two radio lobes 
of the CSO.
Therefore, multi-frequency VLBI observations are needed to detect 
and/or identify the core component to address this issue. 

The case of J1639+1127 also demonstrates that associating the entire 
absorption detected at arcsecond scales through GMRT observations 
with either A or C component would lead to $f_c$ estimates 
that are at least off by a factor of 2. In general, this means that even 
when the core component is accurately identified, $T_S$ measurements 
from DLA studies can have large errors especially if the background 
quasar has significant radio structure at the scales where the 
absorbing gas has strong optical depth gradients. Therefore one 
requires more high-$z$ measurements towards strong core dominated 
sources to address the issues related to the measurements of $T_S$.

Based on the measured E(B-V), the equivalent widths of $\lambda6284$
DIB and Na D lines and various correlations found in our Galaxy we 
predict appreciable column densities of molecular species like
CH, CH$^+$ and CN along this line of sight. Detecting
these species  will allow us to probe the physical and chemical state
of this translucent gas in detail. In addition, this is a good 
target to search for OH lines and molecular lines in radio/mm
wavebands.

\section{acknowledgements}
We thank the referee and Pushpa Khare for useful comments.
We thank GMRT and VLBA staff for their support during the observations. 
The VLBA is run by National Radio Astronomy Observatory. The VLBA data
were correlated using NRAO implementation of the DiFX software
correlator \citep{Deller11} that was developed as part of the Australian Major National
Research facilities Programme and operated under license. The National Radio
Astronomy Observatory is a facility of the National Science Foundation operated
under cooperative agreement by Associated Universities, Inc. GMRT is
run by the National Centre for Radio Astrophysics of the Tata Institute of
Fundamental Research.  We acknowledge the use of SDSS spectra from
the archive (http://www.sdss.org/). R.S.
and P.P.J. gratefully acknowledge support from the Indo-French Centre for the
Promotion of Advanced Research (Centre Franco-Indien pour la promotion de
la recherche avance) under Project N.4304-2.
\def\aj{AJ}%
\def\actaa{Acta Astron.}%
\def\araa{ARA\&A}%
\def\apj{ApJ}%
\def\apjl{ApJ}%
\def\apjs{ApJS}%
\def\ao{Appl.~Opt.}
\def\apss{Ap\&SS}%
\def\aap{A\&A}%
\def\aapr{A\&A~Rev.}%
\def\aaps{A\&AS}%
\def\azh{AZh}%
\def\baas{BAAS}%
\def\bac{Bull. astr. Inst. Czechosl.}%
\def\caa{Chinese Astron. Astrophys.}%
\def\cjaa{Chinese J. Astron. Astrophys.}%
\def\icarus{Icarus}%
\def\jcap{J. Cosmology Astropart. Phys.}%
\def\jrasc{JRASC}%
\def\mnras{MNRAS}%
\def\memras{MmRAS}%
\def\na{New A}%
\def\nar{New A Rev.}%
\def\pasa{PASA}%
\def\pra{Phys.~Rev.~A}%
\def\prb{Phys.~Rev.~B}%
\def\prc{Phys.~Rev.~C}%
\def\prd{Phys.~Rev.~D}%
\def\pre{Phys.~Rev.~E}%
\def\prl{Phys.~Rev.~Lett.}%
\def\pasp{PASP}%
\def\pasj{PASJ}%
\def\qjras{QJRAS}%
\def\rmxaa{Rev. Mexicana Astron. Astrofis.}%
\def\skytel{S\&T}%
\def\solphys{Sol.~Phys.}%
\def\sovast{Soviet~Ast.}%
\def\ssr{Space~Sci.~Rev.}%
\def\zap{ZAp}%
\def\nat{Nature}%
\def\iaucirc{IAU~Circ.}%
\def\aplett{Astrophys.~Lett.}%
\def\apspr{Astrophys.~Space~Phys.~Res.}%
\def\bain{Bull.~Astron.~Inst.~Netherlands}%
\def\fcp{Fund.~Cosmic~Phys.}%
\def\gca{Geochim.~Cosmochim.~Acta}%
\def\grl{Geophys.~Res.~Lett.}%
\def\jcp{J.~Chem.~Phys.}%
\def\jgr{J.~Geophys.~Res.}%
\def\jqsrt{J.~Quant.~Spec.~Radiat.~Transf.}%
\def\memsai{Mem.~Soc.~Astron.~Italiana}%
\def\nphysa{Nucl.~Phys.~A}%
\def\physrep{Phys.~Rep.}%
\def\physscr{Phys.~Scr}%
\def\planss{Planet.~Space~Sci.}%
\def\procspie{Proc.~SPIE}%
\let\astap=\aap
\let\apjlett=\apjl
\let\apjsupp=\apjs
\let\applopt=\ao
\bibliographystyle{mn2e}
\bibliography{mybib}

\end{document}